\documentclass[aps,prd,floatfix,showpacs,superscriptaddress,nofootinbib,11pt]{revtex4-2}

\usepackage{amsmath, amssymb, amsfonts, amsthm, euscript}
\usepackage{graphicx}
\usepackage{xcolor}
\usepackage{hyperref}
\usepackage{bbm}
\usepackage{multirow}
\usepackage{appendix}
\usepackage{subcaption}
\usepackage{tikz}
\usepackage{pgfplots}
\usepackage{braket}
\usepackage{booktabs}
\usepackage{siunitx,pslatex}
\usepackage[export]{adjustbox}
\usepackage{titlesec}  

\hypersetup{
    colorlinks,
    linkcolor={green!50!black},
    citecolor={magenta},
    urlcolor={blue!80!black}
}

\usepackage[top=1.1in, bottom=0.75in, left=1.2in, right=1.2in]{geometry}

\usetikzlibrary{decorations.pathmorphing, decorations.markings, arrows.meta, calc}
\usepgfplotslibrary{fillbetween}

\graphicspath{{Figures/}}

\titleformat{\section}[block]{\normalfont\Large\bfseries\centering}{\thesection}{1em}{}
\titleformat{\subsection}[block]{\normalfont\large\bfseries\centering}{\thesubsection}{1em}{}

\begin{document}

\title{\LARGE Chaos in hyperscaling violating Lifshitz theories}
\author{Nikesh Lilani}\email{imnikeshlilani@gmail.com}\affiliation{Department of Physics and Astronomy, National Institute of Technology Rourkela, Rourkela - 769008, India}

\vspace{0.5cm}

\begin{abstract}
\vspace{0.5cm}
\section*{Abstract}
We holographically study quantum chaos in hyperscaling-violating Lifshitz (HVL) theories (with charge). Specifically, we present a detailed computation of the out-of-time ordered correlator (OTOC) via shockwave analysis in the bulk HVL geometry with a planar horizon topology. We also compute the butterfly velocity ($v_{B}$) using the entanglement wedge reconstruction and find that it matches the result obtained from the shockwave analysis. Using a recently developed thermodynamic dictionary for HVL theories, we express $v_B$ purely in terms of boundary thermodynamic variables. Furthermore, we analyze in detail the behavior of $v_{B}$ with respect to the dynamical critical exponent ($z$), hyperscaling-violating parameter ($\theta$), entropy (more precisely, the ratio of entropy to the central charge, $\tilde{S}$), and charge (more precisely, the ratio of charge to the central charge, $\tilde{Q}$). Interestingly, $v_B$ varies non-monotonically with $z$ for $\tilde{S} < 1$, whereas it increases monotonically with $z$ for $\tilde{S} \geq 1$. Additionally, $v_B$ varies non-monotonically with $\theta$ for non-zero charge. Moreover, $v_B$ monotonically increases with $\tilde{S}$ and decreases with $\tilde{Q}$ for all allowed values of $z$ and $\theta$. All these features are reported for combinations \{$z$, $\theta$, $\tilde{S}$, $\tilde{Q}$\} for which the temperature is positive, the null energy condition is satisfied, and $v_B$ is not superluminal. Unpacking the non-monotonicities in $v_B$ can offer interesting insights into these theories.
\end{abstract} 

\maketitle

\section{Introduction}

Chaos is present in a large class of physical systems that generally exhibit sensitive dependence on initial conditions. Tools of chaos theory are used in a plethora of works ranging from the small quantum domain~\cite{Srednicki:1994mfb} to the large-scale structures in spacetime~\cite{Suzuki:1996gm}. 

Classically, chaos is characterized by an exponential divergence between two phase space trajectories, with the Lyapunov exponent $\lambda_{L}$ parametrizing the rate of this divergence. However, characterization of quantum chaos is a lot more challenging, because the sensitivity in initial conditions cannot be directly computed due to the uncertainty principle. Initially, characterization of quantum chaos was based on comparison between the system's spectrum of energies and the spectrum of random matrices \cite{Ullmo2012INTRODUCTIONTQ}. However, a new approach was developed in \cite{1969JETP...28.1200L} (in the context of semi-classical systems) and \cite{Almheiri:2013hfa}. Quantum chaos can be characterized by the strength of the commutator between two generic operators V and W separated by time t. To be precise, the following quantity is considered: $\left< - [ W(t),V(0)]^{2}\right>_{\beta}$, where expectation value is taken in the thermal state $\beta$. One can also consider the spreading of chaos in spatial directions, by taking the two operators to be also spatially separated. In such a case, the commutator is given by \eqref{eqn:commutator}, where $v_{B}$ is the butterfly velocity. Roughly speaking, $v_{B}$ characterizes the rate at which the information about the applied perturbation "spreads" (or scrambles among the local degrees of freedom) in the system (this notion is made precise in \ref{sec:OTOC}). Focus is laid on $v_{B}$ in this work. 

Gauge/gravity duality  (\cite{tHooft:1993dmi}, \cite{Susskind:1994vu}) is the idea that a gravitational theory in a (d+1) dimensional bulk spacetime is dual to a d dimensional quantum theory (without gravity) residing at the asymptotic boundary of the bulk spacetime. AdS/CFT correspondence \cite{Maldacena:1997re, Witten:1998qj, Gubser:1998bc} is the most well understood realization of this duality. In the framework of AdS/CFT correspondence, a gravitational theory in asymptotically AdS spacetime in the bulk is dual to a conformal field theory (CFT). However, the general principles of gauge/gravity duality can be extended to bulk spacetimes that are not asymptotically AdS and equivalently boudary theories that are not conformally invariant. Lifshitz theories are an example of such theories that are not conformally invariant. These theories exhibit an anisotropic scaling symmetry between space and time, parametrized by the dynamical critical exponent z, i.e. $\{t,x\} \rightarrow \{\alpha^{z}t, \alpha x\}$. Theories with $z \neq 1$ do not support Lorentz invariance and instead possess non-relativistic symmetries. Also, for $z \neq 1$, these theories describe quantum critical systems (non-relativistic). Bulk gravitational theory dual to Lifshitz theory was first proposed in \cite{Kachru_2008}. (see also \cite{Balasubramanian:2008dm, Taylor:2008tg, Ayon-Beato, Mann:2009yx, Bertoldi:2009vn}). The anisotropic scaling symmetry of the boundary theory can be geometrically realized in the bulk via these so-called Lifshitz spacetimes.

Quantum critical point may be characterized by several other critical exponents that satisfy certain relations and thereby are dependent on each other. Hyperscaling relations are a certain class of critical exponent relations in which there is an explicit appearance of dimensionality of the space \cite{Widom1965SurfaceTA}. There are critical theories that violate this hyperscaling relation. For such theories the heat capacity $C_{v} \sim T^{(d-1-\theta )/z}$ (\cite{Gouteraux:2011ce},\cite{Huijse:2011ef}), where $\theta$ is called the hyperscaling-violating parameter. Essentially, $\theta$ reduces the effective dimensionality of the theory.
For hyperscaling-violating and Lifshitz (HVL) theoreies, the dual bulk black hole solutions have been obtained in several works \cite{Alishahiha:2012qu,Dong:2012se,Gouteraux:2012yr,Gath,Bueno,Pedraza:2018eey}. In this paper, we closely follow the discussion in \cite{Pedraza:2018eey}. 

In recent years, the study of chaotic dynamics of strongly-coupled many-body quantum systems via gauge-gravity  duality has witnessed a significant interest \cite{Shenker_2014,Roberts_2015,PhysRevLett.117.091602,Perlmutter:2016pkf}. For a comprehensive review of holographic chaos, refer \cite{Jahnke:2018off}. Through \cite{Sekino_2008}, it was established that black holes are the fastest scramblers in nature. Thus, in the context of AdS/CFT, the rapid thermalization of a local perturbation in the boundary CFT is understood through the fast scrambling dynamics of a black hole in the bulk. It is also known that black holes in holography can be characterized via quantum chaos \cite{Sekino_2008}, \cite{Maldacena:2001kr}. Possessing maximal chaos is thought of as a criterion for a QFT to have a gravity dual \cite{Maldacena:1997re}, \cite{Shenker_2014}, \cite{Shenker:2014cwa}.

In this work, we holographically study quantum chaos in HVL theories (with charge). In particular, we compute the out-of-time-ordered correlator (OTOC) (we make the notion precise in \ref{sec:OTOC}) via shockwave analysis in the bulk geometry (for planar horizon topology) and thereby compute the various chaos parameters, namely the Lyapunov exponent ($\lambda_{L}$), butterfly velocity ($v_{B}$) and the scrambling time ($t_{*}$) for the HVL theories. One of the motivations for studying OTOCs in HVL theories is the recent proposals for experimentally measuring OTOCs in quantum systems \cite{PhysRevA.94.040302,PhysRevA.94.062329,PhysRevX.7.031011,Yao:2016ayk}. We also compute $v_{B}$ using the entanglement wedge reconstruction and find the results obtained from both the methods, to match. Furthermore, we study how boundary thermodynamics and theory parameters (z and $\theta$) affect information scrambling. To achieve this, we use a recently developed thermodynamic dictionary for the HVL theories \cite{Cong:2024pvs} to express $v_B$ in terms of boundary thermodynamic variables. We study the variation of $v_B$ with respect to the theory parameters $z$ and $\theta$, as well as the boundary thermodynamic variables: entropy over central charge ($\tilde{S} \sim S/C$) and charge over central charge ($\tilde{Q} \sim Q/C$). Several interesting features are obtained.
 
The structure of this paper is as follows: In \ref{sec: background}, we discuss the charged hyperscaling-violating Lifshitz black hole solution and also discuss the constraints on z and $\theta$ that arise from the null-energy condition. In \ref{sec:OTOC}, we provide a detailed analysis of computing the OTOC for HVL theories via the shockwave analysis. In \ref{sec: EW}, we compute $v_{B}$ using the entanglement wedge reconstruction and find the results of \ref{sec:OTOC} and \ref{sec: EW} to match. In \ref{sec: vb_bound_quant}, we discuss the thermodynamic dictionary and express $v_B$ in terms of the boundary thermodynamic variables. In \ref{sec: variation}, we discuss the variation of $v_{B}$ with respect to the various z, $\theta$, $\tilde{S}$, and $\tilde{Q}$. We conclude in \ref{sec: conclusion}.

\section{Background}
\label{sec: background}
In this paper, we follow the discussion in \cite{Pedraza:2018eey}. The HVL black hole solution obtained in \cite{Pedraza:2018eey} is a generalization of charged black brane solutions with arbitrary z and $\theta $ (of \cite{Alishahiha:2012qu}) to other topologies (namely spherical and hyperbolic). 

The line element is given as follows
\begin{equation}
    \label{eqn:HVLmetric}
    ds^{2} = \left(\frac{r}{r_{F}}\right)^{\frac{-2 \theta }{d-1}} \left[ -\left(\frac{r}{L}\right)^{2 z} f(r) dt^{2} + \frac{L^{2}}{f(r) r^{2}} dr^{2} + r^{2} d\Omega^{2}_{k,d-1}\right]
\end{equation}
where f(r) is the blackening function given as
\begin{equation}
    \label{eqn: blackening_HVL}
    f(r) = 1 + k \frac{(d-2)^{2}}{(d-\theta + z-3)^{2}} \frac{L^{2}}{r^{2}}-\left(\frac{r_{h}}{r}\right)^{d-\theta + z-1}\left[1+ k \frac{(d-2)^{2}}{(d-\theta + z-3)^{2}} \frac{L^{2}}{r_{h}^{2}}  + \frac{q^{2}}{r_{h}^{2(d-\theta + z-2)}}\right] + \frac{q^{2}}{r^{2(d-\theta+z-2)}}
\end{equation}
$r_h$ is the horizon radius, d is the number of boundary dimensions, q is the charge parameter, k is the factor that characterizes the topology of the horizon ($ k = \{-1, 0, 1\}$ for hyperbolic, planar and spherical horizon topology respectively), $r_{F}$ is the large radius upto which the black hole geometry is considered to be valid, ($r_{F}$ corresponds to the UV cutoff scale in the boundary), and L is the bulk curvature radius. The temperature can be computed from the blackening function.
\begin{equation}
    \label{eqn:temp_HVL}
    T = \frac{f'(r_h)}{4\pi} \left( \frac{r_h}{L} \right)^{z+1} = \frac{r_h^z}{4\pi L^{z+1}} \left[ d - \theta + z - 1 + \frac{k(d-2)^2}{d - \theta + z - 3} \frac{L^2}{r_h^2} - \frac{(d - \theta + z - 3) q^2}{ r_h^{2(d - \theta + z - 2)}} \right]
\end{equation}
The charged and static black hole solutions  given by \eqref{eqn:HVLmetric} were analytically constructed from the following generalized Einstein-Maxwell-Dilaton (EMD) action 
\begin{equation}
    \label{EMD_action}
    S = -\frac{1}{16 \pi G} \int d^{d+1}x \sqrt{-g} \left[ R - \frac{1}{2} (\nabla \phi)^2 + V(\phi) 
- \frac{1}{4} X(\phi) F^2 - \frac{1}{4} Y(\phi) H^2 - \frac{1}{4} Z(\phi) K^2 \right]
\end{equation}
Here F, H and K are the field strengths corresponding to the three Abelian gauge fields A,B,C, given as $F = dA$, $H= dB$ and $K = dC$. $\phi$ is a real scalar field known as the dilaton field. X, Y and Z are the respective coupling functions of the gaugle fields and the dilaton field given as
\begin{equation}
    \label{X,Y,Z}
    X = X_{o} e^{\lambda_{1}\phi}, \hspace{0.3cm} Y = Y_{o} e^{\lambda_{2}\phi}, \hspace{0.3cm} Z = Z_{o} e^{\lambda_{3}\phi} 
\end{equation}
Here, $X_{o}$, $Y_{o}$, and $Z_{o}$ are positive parameters that quantify the strength of coupling between gravity and the three gauge fields respectively, and $\lambda_{i}s$ are constants given as 
\begin{equation}
    \label{eqn: lambdas}
    \lambda_{1} = \frac{-2(d-1-\theta + \theta/(d-1)) }{\tilde{\gamma}}, \hspace{0.3cm} \lambda_{2} = \frac{-2(d-2)(d-\theta -1)}{\tilde{\gamma} (d-1)}, \hspace{0.3cm} \lambda_3 = \frac{\tilde{\gamma}}{d-\theta -1}
\end{equation}
where $\tilde{\gamma}$ is given as 
\begin{equation}
    \label{eqn: gamma}
    \tilde{\gamma} = \sqrt{2 (d-\theta -1)\left(z-1 -\frac{\theta}{d-1}\right)}
\end{equation}
Note that the gauge field A supports the Lifshitz asymptotics, B supports the other non-planar horizon topologies (hyperbolic and spherical) and C allows for a non-zero electric charge. In this work we deal only with planar horizon topology ($k=0$) and set $L=1$. For $k=0$ and $L=1$, f(r) and temperature are given as
\begin{equation}
    \label{eqn:f(r)_k=0}
    f(r) = 1 -\left(\frac{r_{h}}{r}\right)^{d-\theta + z-1}\left[1 + \frac{q^{2}}{r_{h}^{2(d-\theta + z-2)}}\right] + \frac{q^{2}}{r^{2(d-\theta+z-2)}}
\end{equation}
\begin{equation}
    \label{eqn:temp_k=0}
    T = \frac{r_h^z}{4\pi } \left[ d - \theta + z - 1  - \frac{(d - \theta + z - 3) q^2}{ r_h^{2(d - \theta + z - 2)}} \right]
\end{equation}

\subsection{Constraints on z and $\theta$}
\label{subsec:constraints}
It is assumed that the null energy condition serves as a sufficient condition to have a physically consistent holographic dual in the semi-classical limit. The requirement of satisfying the null energy condition imposes certain constraints on z and $\theta$. The null energy condition is given as 
\begin{equation}
    \label{eqn:NEC}
    T_{\mu\nu}\xi^{\mu}\xi^{\nu} \geq 0
\end{equation}
or equivalently,
\begin{equation}
    \label{eqn:NEC_2}
    R_{\mu\nu}\xi^{\mu}\xi^{\nu} \geq 0
\end{equation}
where $\xi^{\mu}$ is a null vector. Considering two orthogonal null vectors, we obtain the following two inequalities from \eqref{eqn:NEC_2} (for $k=0$)
\begin{equation}
    \label{eqn: ineq_1}
    (d - 1-\theta)((d-1)(z - 1) - \theta) \geq 0
\end{equation}
\begin{equation}
    \label{eqn: ineq_2}
    \frac{r^2}{ (z - 1)(d - 1-\theta + z)} 
+ \frac{q^2 (d - 1- \theta)((d-1)(z - 1) - \theta)}{r^{2(d-1 - \theta + z - 2)}} \geq 0
\end{equation}
Note that the black hole solution is only valid for $\theta < d-1$ and $d -\theta + z -3 > 0$. Combining these inequalities with \eqref{eqn: ineq_1} and \eqref{eqn: ineq_2}, we get the following constraints on the values of z and $\theta$ for $k=0$.
\begin{equation}
\begin{aligned}
    z < 1 &\implies \text{no solution} \\[0.3cm]
    1 \leq z < 2 &\implies \text{solution exists for } \theta \leq (d-1)(z-1) \\[0.3cm]
    z \geq 2 &\implies \text{solution exists for } \theta < d-1
\end{aligned}
\end{equation}
These constraints play an important role in the discussion in \ref{sec: variation}.

\section{Out-of-time ordered correlator (OTOC)}

\label{sec:OTOC}
For two spatially separated generic Hermitian operators V and W, quantum chaos can be characterized by the following commutator $[ W(0,-t),V(x,0)]$. Note that, since it is of interest to study thermal systems, an expectation value of the commutator is taken in a thermal state $\beta$. However, the thermal expectation value may attain a negative sign which can thereby lead to cancellations. To overcome this, the commutator is squared and thus the precise quantity to be considered, is the following.
\begin{equation}
 \label{eqn:commutator}
    C(x,t) = \left< - [ W(0,-t),V(x,0)]^{2}\right>_{\beta}
\end{equation}
where the overall minus sign is introduced to make C(x,t) positive. C(x,t) characterizes the strength of the effect that the perturbation $W$ at an earlier time -t and $x=0$ has, on the measurement of the operator V at $t=0$ and at a spatial point x. We choose to work with past time (-t) because it is convenient when we consider the holographic picture. \eqref{eqn:commutator} can be expanded as follows.
\begin{equation}
  \label{eqn:otoc}
    C(x,t) = 2 - \left < W(0,-t) V(x,0)W(0,-t) V(x,0)\right>
\end{equation}
The second term in \eqref{eqn:otoc} is called the out-of-time ordered correlator (OTOC) (simply because the operators are not time-ordered). The vanishing of the OTOC is considered as a diagnostic of chaos in a physical system. This can be realized by considering the following two states. 
\begin{equation}
    \ket{\psi_{1}} = W(0,-t)V(x,0)\ket{\beta}
\end{equation}
\begin{equation}
    \ket{\psi_{2}} = V(x,0)W(0,-t)\ket{\beta}
\end{equation}
OTOC is the overlap between the above two states. The state $\ket{\psi_1}$ can be physically described as follows: at time $t = 0$ (and at spatial point $x$), the state $V(x,0)\ket{\beta}$ is prepared. The state is then evolved backward in time, a small perturbation $W$ is applied, and the state is evolved forward in time again. If the perturbation $W$ is applied sufficiently early in the past, and the system is chaotic (i.e., highly sensitive to initial conditions), the operator $V$ will fail to reappear at $t = 0$.
On the other hand, the state $\ket{\psi_2}$ is understood as follows: a perturbation $W$ is applied at time $-t$, and then the system is evolved in such a way that the operator $V$ appears at $t = 0$. For chaotic systems, the overlap between the two states $\ket{\psi_1}$ and $\ket{\psi_2}$ vanishes. As a result, OTOC vanishes, leading to the growth of the commutator, as is evident from equation \eqref{eqn:otoc}. In the case of conformal field theories (CFTs) and higher-dimensional SYK models, this commutator between two spatially separated operators shows exponential growth.
\begin{equation}
    \label{eqn:exp_commutator}
    C(x,t) \sim \exp{\left[\lambda_{L}\left(t - t_{*}-\frac{|x|}{v_{B}}\right)\right]}
\end{equation}
Here, $\lambda_{L}$ is the quantum Lyapunaov exponent, $t_{*}$ is the scrambling time and $v_{B}$ is the butterfly velocity. The scrambling time $t_{*}$ is defined as the time at which the commutator with $x=0$ becomes order unity. In chaotic systems, the commutator with $x=0$ exhibits an exponential growth upto the scrambling time. After the scrambling time, the information about a local perturbation starts scrambling among the local degrees of freedom at a constant rate, characterized by the butterfly velocity $v_{B}$. After time t of inserting the perturbation operator W ($|t| > |t_{*}|$), the commutator is order unity in the region given by ($ |x| < v_{B}|t-t_{*}|$), and this region is said to be the "size" of the operator W (it is the region in which the measurement of any operator is significantly affected by W). 

\subsection{Kruskal extension and the shockwave analysis}

In this subsection, we compute the various chaos parameters for HVL theories via the shockwave analysis in the bulk.
It is standard to work with the two-sided eternal black hole geometry, which corresponds to a thermo-field double state (denoted as $\ket{TFD}$) in the boundary. Note that we will consider planar horizon topology (i.e. $k=0$) for our analysis throughout.

In order to work in the two-sided geometry, the metric \eqref{eqn:HVLmetric} is written in the Kruskal co-ordinates. The Kruskal co-ordinate transformation is given as 
\begin{equation}
\label{eqn: kruskal transformation}
    u = exp\left[\frac{2\pi}{\beta}(r_{*}-t)\right],
  \hspace{0.4cm}  
  v = -exp\left[\frac{2\pi}{\beta}(r_{*}+t)\right]
\end{equation}
Where, $\beta$ is the inverse Hawking temperature and $r_{*}$ is the tortoise co-ordinate.

\begin{equation}
\label{eqn:tortoiseHVL}
    dr_{*} = \frac{dr (r^{z+1})}{f(r)}
\end{equation}
In the Kruskal co-ordinates the metric takes the following form 
\begin{equation}
    \label{eqn:kruskal_metric}
    ds^{2} = 2A(u,v)dudv + B(u,v)dx^{i}dx^{i}
\end{equation}
where A(u,v) and B(u,v) are related to the blackening function as follows.
\begin{equation}
    \label{eqn: relation of A and B}
    A(u,v) = \frac{f[r(u,v)]\left[r(u,v)\right]^{2\left(z-\frac{\theta}{d-1}\right)}}{(2 \alpha^{2} uv) r_{F}^{\frac{-2 \theta}{d-1}}}; \hspace{0.3cm} B(u,v) = \frac{[r(u,v)]^{2\left(1-\frac{\theta}{d-1}\right)}}{r_{F}^{\frac{-2\theta}{d-1}}}    
\end{equation}

Here $\alpha = \frac{2 \pi}{\beta}$. In order to know the explicit expressions of A and B in terms of u and v, we need to find $r[u,v]$. \eqref{eqn:tortoiseHVL} is not integrable. However, as we shall discuss, the shockwave is localized at the horizon and thus we deal with the near-horizon limit. The integral given by \eqref{eqn:tortoiseHVL}, can be expanded around the horizon. The dominating term in the expansion is the logarithmic term given as.
\begin{equation}
    \label{eqn: log_term}
    r_{*} \approx \frac{1}{f'(r_{h})r_{h}^{z+1}} \log(r-r_{h})
\end{equation}
Multiply u and v, given by \eqref{eqn: kruskal transformation}, we get the following.
\begin{equation}
    \label{eqn: uv}
    uv = - e^{2\alpha r_{*}} = -e^{4 \pi T z_{*}}
\end{equation}
Substituting \eqref{eqn: log_term}, \eqref{eqn:temp_HVL} in \eqref{eqn: uv}, we get the following.
\begin{equation}
    \label{eqn: z=1-uv}
    r = r_{h} - uv
\end{equation}
Putting $u=0$ or $v=0$, we get that $r=r_{h}$. This implies that $u=0$ and $v=0$ are the two horizons of the two-sided geometry (as is the case in Kruskal co-ordinates). Substituting \eqref{eqn: z=1-uv} in \eqref{eqn: relation of A and B}, we get the explicit form of A and B in terms of the Kruskal co-ordinates (u and v).

\begin{figure}[h!]
    \centering
    \begin{tikzpicture}[scale=1.1] 

    \draw[ultra thick] (-3.0,-3.1) -- (-3.0,3.1); 
    \draw[ultra thick] (3.0,-3.1) -- (3.0,3.1);   

    \draw[thick,decorate,decoration={zigzag,amplitude=2pt,segment length=4pt}] 
        (-3,3) .. controls (-1.5,2.5) and (1.5,2.5) .. (3,3); 
    \draw[thick,decorate,decoration={zigzag,amplitude=2pt,segment length=4pt}] 
        (-3,-3) .. controls (-1.5,-2.5) and (1.5,-2.5) .. (3,-3); 

    \draw[thick] (-3,3) -- (3,-3);
    \draw[thick] (-3,-3) -- (3,3);

    \draw[thick,red] (-2.4,2.9) -- (3,-2.5);

    \draw[thick,blue] (1.1,-0.6) -- (3.0,1.4);
    \draw[thick,blue,dashed] (1.89,-1.28) -- (3.0,-0.1);
    \draw[thick,blue] (0.55,-2.58) -- (1.89,-1.28);

    \node[rotate=45] at (2.0,1.7) {$u \, (v = 0)$};
    \node[rotate=-45] at (-2.0,1.7) {$v \, (u = 0)$};

    \draw[<->,black] (2.1,-1.1) -- (1.38,-0.36);
    \node[rotate=30,black] at (2.1,-0.4) {$h(x)$};

    \end{tikzpicture}
    \captionsetup{justification=raggedright}
    \caption{The Penrose diagram, explaining the effect of the shockwave created near the horizon. The red line represents the shockwave created by the perturbation particle W, which is inserted at a sufficiently earlier time. In the absence of the shockwave, particle V would have emerged from the past interior and would have reached the boundary at $t=0$. However, due to the shockwave, the V particle suffers a shift in its trajectory, which is parametrized by $h(x)$. This causes a delay in the appearance of operator V at the boundary. If the W perturbation is applied early enough (i.e., $ |t| \gtrsim |t_{*}| $), the trajectory of the particle gets shifted to the point that it gets engulfed by the future interior and the V particle fails to reach the boundary altogether.}
    \label{fig:OTOC}
\end{figure}
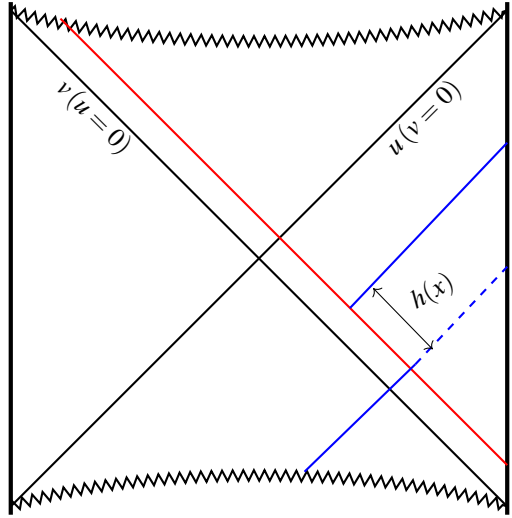

Now, we compute the OTOC. 
In the framework of AdS/CFT, each operator $\hat{Q}$ in the boundary corresponds to a scalar perturbation $\phi(x,t)$ in the bulk. Consider acting an operator W(0,-t) on $\ket{TFD}$ in the right boundary. In the two-sided black hole geometry, this corresponds to a particle coming out from the past interior, reaching the boundary at time -t and then falling towards the future interior. The energy of the particle falling into the black hole increases exponentially and it depends on the temperature of the black hole.
\begin{equation}
    \label{eqn:exp_eng}
    E = E_{o}e^{\frac{2\pi}{\beta}t}
\end{equation}
Here $E_{o}$ is the energy of the particle when it's near the boundary. Since, the energy of the particle gets exponentially blue-shifted, a sufficiently earlier perturbation in the boundary, W(0,-t), leads to a non-trivial modification of the bulk geometry. The large energy of the particle leads to a back reaction in the geometry, which simply corresponds to a shockwave geometry \cite{Roberts_2015}. The energy distribution of this perturbation is compressed along u and stretched along v. Thus, for sufficiently large time $|t|$, the perturbation gets localized along the horizon $u = 0$ (refer figure \ref{fig:OTOC}) and, the stress energy tensor attains the following form.
\begin{equation}
\label{eqn:T_shock}
    T_{uu}^{shock} = E_{o}e^{\frac{2\pi}{\beta}t}\delta(u)a(x)
\end{equation}
Where a(x) is some function to be determined. The effect of this shock, localized along the horizon $u=0$ is that the particle coming from the past interior suffers a shift in the trajectory (refer figure), which is given as
\begin{equation}
    \label{eqn:shift}
    v \rightarrow v + \Theta(u) h(x), \hspace{0.5cm} u \rightarrow u
\end{equation}
Here, the step function $\Theta(u)$ accounts for the fact that the causal future of the particle is affected after it encounters the horizon $u=0$ (and thus there's no affect on the causal past). h(x,t) is a function that characterizes the amount of shift and it is to be determined by the Einstein equation. To understand the relation between this function h(x,t) and the commutator (and thereby the OTOC), consider the bulk picture of the two states $\ket{\psi_{1}}$ and $\ket{\psi_{2}}$. The bulk picture of $\ket{\psi_{1}}$ is as follows: Initially,the system is prepared in such a way that the V particle from the past interior reaches the boundary at $t=0$. However, the insertion of the operator W at a sufficiently earlier time makes the particle encounter a localized shock at the horizon, which results in the shift of particle's trajectory, which in turn, causes a time delay in the appearance of the operator V at the boundary. On the other hand, the state $\ket{\psi_{2}}$ corresponds to the V particle reaching the boundary at $t=0$, "after" suffering the shift due to the shock. Thus, the overlap between the two states (and thus the OTOC) keeps decreasing for earlier and earlier times. 
After a certain past time, the shift suffered by the particle is such that the particle gets trapped inside the future interior and then it cannot escape to the boundary, making the overlap and thus the OTOC completely vanish. This time is given by the scrambling time, $|t| \gtrsim |t_{*}|$. 

The shift in the co-ordinates \eqref{eqn:shift} results in the following modified metric.
\begin{equation}
\label{eqn:modified_metric}
    ds^{2} = 2A(u,v)dudv + B(u,v)dx^{2} - 2A(u,v)h(x,t)\delta(u)du^{2}
\end{equation}
The total stress tensor is now given as
\begin{equation}
    \label{eqn:total_T}
    T = T_{o} + T^{shock}
\end{equation}
Here, $T_{o}$ is the initial unperturbed stress tensor and the only non-zero component of $T^{shock}$ is given by \eqref{eqn:T_shock}. Substituting \eqref{eqn:total_T} and \eqref{eqn:modified_metric} in the Einstein equation and solving for the uu component, we arrive at the following.
\begin{equation}
    \label{eqn:ode_for_h}
    \left(\partial_{i}\partial_{i} - \frac{d-1}{2}\frac{\partial_{u}\partial_{v}B}{A}\right) h(x)\delta(u) = \frac{8 \pi G_{N} E_{o} B}{A} e^{\frac{2\pi t}{\beta}}\delta(u)a(x)
\end{equation}
For large x, ($|x| >> 1)$, the function a(x) can be replaced by the Dirac delta function , because the solution depends only on the integral of a(x). Solving for large x, we get the following solution for h(x)
\begin{equation}
    \label{eqn: h(x,t)}
    h(x) = \frac{e^{\frac{2\pi}{\beta}(t-t_{*})- \chi |x|}}{|x|^{\frac{d-2}{2}}}
\end{equation}
Where $\chi$ and $t_{*}$ are given as
\begin{equation}
    \label{eqn: chi}
    \chi = \sqrt{\frac{d-1}{2}\left(\frac{\partial_{u}\partial_{v}B(0)}{A(0)}\right)}
\end{equation}
\begin{equation}
    \label{eqn: tstar}
    t_{*} = \frac{2 \pi}{\beta} \log\left(\frac{A(0)}{8 \pi G_{N}E_{o}B(0)}\right)
\end{equation}
where $\partial_{u}\partial_{v}B(0)$ denotes the double derivative of B evaluated at the horizon and A(0), B(0) denote the functions A and B, evaluated at the horizon. Comparing \eqref{eqn: h(x,t)} with \eqref{eqn:exp_commutator}, we can read off the butterfly velocity and the Lyapunov exponent.
\begin{equation}
    \label{eqn: VBOTOC}
    v_{B} = \frac{2\pi}{\beta \chi}
\end{equation}
\begin{equation}
    \label{eqn:lyapunov}
    \lambda_{L} = 2\pi T 
\end{equation}
We can compute $v_{B}$ and $t_{*}$ by evaluating the values of A, B and $\partial_{u}\partial_{v}B$ at the horizon $u=0$. Note that, at the horizon, A attains a $\frac{0}{0}$ form. Thus we compute the limit of A as $u \rightarrow 0$, which is well defined. The required horizon-limit values are
\begin{equation}
    \label{eqn:A(0)}
    A(0) = \lim_{u \to 0}  A = \frac{r_h^{-1 - 2d} \left(\frac{r_h}{r_F}\right)^{-\frac{2 \theta}{d-1}} \left(-r_h^{2(d + z)}(d-1 + z - \theta) - q^2 r_h^{4 + 2\theta} (3 - d - z + \theta)\right)}{2 \alpha^2}
\end{equation}
\begin{equation}
    \label{eqn:B(0)}
    B(0) = \left(\frac{r_h}{r_F}\right)^{-\frac{2 \theta}{d - 1}} r_h^2
\end{equation}
\begin{equation}
    \label{eqn:deludelv}
    \partial_{u}\partial_{v}B(0) = \frac{-2 r_h \left(\frac{r_h}{r_F}\right)^{-\frac{2 \theta}{d-1}} + 2 r_h^2 \left(\frac{r_h}{r_F}\right)^{-1 - \frac{2 \theta}{d-1}} \theta}{(d-1) \, r_F}
\end{equation}
Using \eqref{eqn:A(0)}, \eqref{eqn:B(0)}, \eqref{eqn:deludelv}, and \eqref{eqn:temp_HVL} (for $k=0$) we can compute the three chaos parameters for the HVL theories. .

\begin{equation}
  \begin{split}
    \label{tstar_HVL}
    t_{*} = & \, \frac{1}{2} r_h^{1 + z} \left(-2 q^2 r_h^{3 - 2d - 2z + 2\theta} (d -2 + z - \theta) + (1 + q^2) \left(\frac{1}{r_h}\right)^{d + z - \theta} ( d - 1 + z - \theta)\right) \\
    & \times \log \left( \frac{ \left(r_h^{2(d + z)}(1 - d - z + \theta) + q^2 r_h^{4 + 2\theta} ( d -3 + z - \theta)\right)}{4 G_{N} \pi E_{o} r_{h}^{5} \left(-2 q^2 r_h^{3 - d - z + 2\theta}(d -2 + z - \theta) + (1 + q^2) r_h^{\theta}(d -1 + z - \theta)\right)^2} \right)
  \end{split}
\end{equation}

\vspace{0.5cm}
\begin{equation}
   \label{eqn: vB_HVL}
   v_{B}^{2} = \frac{ r_h^{2(z-1)} (d -1 + z - \theta) + q^2 r_h^{-2(d-1- \theta)} (3 - d - z + \theta)}{2 (d-1 - \theta)}
\end{equation}
\begin{equation}
    \label{eqn: lyapunov}
    \lambda_{L} = \frac{r_h^z}{2} \left[ d - \theta + z - 1 - \frac{(d - \theta + z - 3) q^2}{ r_h^{2(d - \theta + z - 2)}} \right]
\end{equation}
For $r_{h} = 1$, the expressions for these parameters become simpler. 
\begin{equation}
    t_{*} = \frac{1}{2} \left( d -1 + z - \theta + q^2 (3 -d- z + \theta)\right) \log \left( \frac{1}{4 G_{N} \pi E_{o} \left(1 - d - z + \theta - q^2 (3 - d - z + \theta)\right)} \right)
\end{equation}
\begin{equation}
    \label{eqn:VB_RH=1}
    v_{B}^{2} = \frac{d-1 + z - \theta + q^2 (3 - z + \theta-d)}{2 (d-1 - \theta)}
\end{equation}
\begin{equation}
    \lambda_{L} = \frac{1}{2} \left(-1 + d + z - \theta + q^2 (3 -d - z + \theta)\right)
\end{equation}
Note that for $\theta = 0$, $r_h=1$, and $q=0$, the result (\ref{eqn: vB_HVL}) matches the result obtained in \cite{Baishya:2024sym}. Also, as an important sidenote, we point out that the butterfly velocity \eqref{eqn: vB_HVL} is equal to the velocity of entanglement growth during thermalization at the saturation point \footnote{Author would like to thank 
Mohammad Reza Mohammadi Mozaffar for pointing this out.} (for $q = 0$) \cite{Alishahiha:2014cwa} , suggesting an interesting connection between the two notions. 

\section{Entanglement wedge}

\label{sec: EW}
In this section, we compute $v_{B}$ using another method, namely the entanglement wedge method. Through the works \cite{Czech:2012bh, Wall:2012uf, Headrick:2014cta, Dong:2016eik}, it was established that a certain sub-region A on the boundary is dual to the entanglement wedge in the bulk. The entanglement wedge is the region in the bulk, bounded by the sub-region A and the extremal surface homologous to A. In \cite{Mezei:2016wfz}, it was shown that the butterfly velocity (for isotropic and planar bulk geometries) can be computed using the entanglement wedge reconstruction. 

\begin{figure}[h!]
    \centering
    \begin{tikzpicture}

    \definecolor{yellowfill}{RGB}{255, 255, 153}
    \definecolor{reddot}{RGB}{255, 0, 0}

    \draw[thick] (-5,2.6) -- (0,2.6) node[midway, above] {Horizon};
    \draw[thick] (-5,0) -- (0,0) node[midway, below] {Boundary};

    \draw[thick] (2,2.6) -- (7,2.6) node[midway, above] {Horizon};
    \draw[thick] (2,0) -- (7,0) node[midway, below] {Boundary};

    \fill[yellowfill] (-1,0) arc[start angle=0, end angle=180, radius=1.2] -- cycle;
    \draw[thick] (-1,0) arc[start angle=0, end angle=180, radius=1.2];

    \fill[reddot] (-2.17,1.11) circle (0.1);

    \coordinate (A) at (3,0);
    \coordinate (B) at (3,2.05);
    \coordinate (C) at (6,2.05);
    \coordinate (D) at (6,0);
    \draw[ultra thick] (A) to[out=90, in=90, looseness=0.5] (B) to[out=100, in=90, looseness=0.5] (C) to[out=90, in=90, looseness=0.5] (D);
    \fill[yellowfill] (A) to[out=90, in=90, looseness=0.5] (B) to[out=90, in=90, looseness=0.5] (C) to[out=90, in=90, looseness=0.5] (D);

    \fill[reddot] (4.5,2.39) circle (0.1);

    \draw [ultra thick,blue, arrows = {-Stealth}] (6.2,0.6) -- (6.8,0.6) node[right] {$v_B$};
    
    \end{tikzpicture}
    \captionsetup{justification=raggedright}
    \caption{The growth of the entanglement wedge (shaded in yellow) as the particle, originating from the boundary (represented by the red dot), propagates towards the horizon. The figure on the left shows the entanglement wedge at some time $t$ and the figure on the right shows the entanglement wedge at a sufficiently later time $t'$, when the RT surface reaches near the horizon and the near-horizon profile of the RT surface is given by $u(\tilde{r})$.}
    \label{fig:EW}
\end{figure}
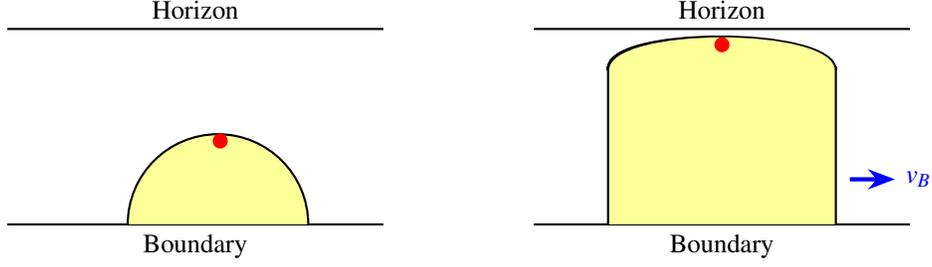

Consider some perturbation in the boundary theory. As time passes, the information from the perturbation gets delocalized over a larger and larger region, i.e. the size of the applied operator increases. The size of the operator is the smallest region that contains the information of the applied perturbation operator. In the dual bulk picture, it corresponds to the extremal surface "just" enclosing the particle (i.e. the particle is at the tip of the extremal surface at all times) (refer figure \ref{fig:EW}). It's important to note that the background geometry is static and it's only the position of the particle that is time-dependent. So the RT surface~\cite{Ryu_2006} is used as the extremal surface and not the HRT surface~\cite{Hubeny:2007xt}. Thus, in this case the entanglement wedge is a constant t hyper-surface, bounded by the RT surface and the boundary sub-region. 

The area functional (or equivalently, the entanglement entropy functional) is given as follows.
\begin{equation}
    \label{eqn: SEE}
    S_{EE} = 2\pi \int \sqrt{\gamma}
    \, d^{d-1}\xi
\end{equation}
Here $\gamma $ is the determinant of the induced metric on the co-dimension 2 surface in the bulk and $\xi$ are the co-ordinates on the surface. Extremising the above action (or equivalently the induced metric) will give rise to the equation of the RT surface. So we first compute the induced metric on the surface. Since the surface is embedded in a $t= constant$ hypersurface, the tt component of the induced metric vanishes. Further, taking $\tilde{r}$ to be the radial co-ordinate in the $x^{i}$ direction  ($\tilde{r} = |x|$) and parametrizing r with $\tilde{r}$ ($r = r(\tilde{r})$), we get the following induced metric from \eqref{eqn:HVLmetric} 
\begin{equation}
    \label{eqn: induced metric}
    \gamma_{ab}d\xi^{a}d\xi^{b} = \left[\frac{(z')^{2}\left(\frac{r}{r_{F}}\right)^{\frac{-2\theta}{d-1}}}{r^{2}f(r)} + \left(\frac{r}{r_F}\right)^{-\frac{2 \theta}{d - 1}} r^2
\right]d\tilde{r}^{2} + \left[\left(\frac{r}{r_F}\right)^{-\frac{2 \theta}{d - 1}} r^2
\right]\tilde{r}^{2}d\Omega_{0,d-2}^{2}
\end{equation}
In order to compute the butterfly velocity, a near-horizon analysis of the RT surface is done (refer figure \ref{fig:EW}). Near the horizon, the following form of r($\tilde{r}$) can be taken.
\begin{equation}
    \label{eqn: z(r)} 
    r (\tilde{r}) = 1 - \epsilon u(\tilde{r})^{2}
\end{equation}
For simplicity, here we have taken the horizon radius $r_{h} = 1$, $\epsilon$ is some arbitrarily small parameter and u($\tilde{r}$) is the function to be determined by extremizing \eqref{eqn: SEE}. Substituting \eqref{eqn: z(r)} in the induced metric and expanding near the horizon, upto $\mathcal{O}(\epsilon)$, we get the following.
\begin{equation}
    \label{eqn: det_induced}
    \sqrt{\gamma} = r_{F}^{\theta} \tilde{r}^{d-2}\left(1-\epsilon\frac{\left(4 r_{F}^{\frac{2 \theta}{d-1}} (u')^{2} + (d-1)u^{2}f'_{o}\left(2r_{F}^{\frac{2 \theta}{d-1}}\left[1-\frac{\theta}{d-1}\right]\right)\right)}{2 r_{F}^{\frac{2 \theta}{d-1}} f'_{o}}\right)
\end{equation}
Here $f_{o}'$ represents the derivative of the blackening function, evaluated at the horizon. 
Extremising \eqref{eqn: det_induced} (and thereby $S_{EE}$) we arrive at the following differential equation in u($\tilde{r}$).
\begin{equation}
\label{eqn: ODE for u}
    u''(\tilde{r}) + (d-2) \frac{u'(\tilde{r})}{\tilde{r}} - \frac{f_{o}'}{2}(d-1-\theta) u(\tilde{r}) = 0
\end{equation}
The RT surface starts to depart the near-horizon region for large $\tilde{r}$. To put it precisely, the RT surface stays close to the horizon up to the point where $\epsilon u(\tilde{r})^{2} \sim$ $\mathcal{O} (1)$, and after that, the surface departs the near horizon region and reaches the boundary to order one distance. As suggested in \cite{Dong_2022}, we can encapsulate this behaviour via the following ansatz.
\begin{equation}
    \label{eqn: ansatz}
    u(\tilde{r}) \sim \frac{e^{\mu \tilde{r}}}{\tilde{r}^{n}}
\end{equation}
where n is some positive integer. Also, the particle touches the tip of the RT surface at all times. Hence, taking the tip of the RT surface to be the origin, we can set $u(0,t) \sim e^{-\frac{2 \pi t}{\beta}}$. Thus u($\tilde{r}$,t) is given as follows.
\begin{equation}
    \label{eqn: u(r,t)}
    u(\tilde{r},t) \sim \frac{e^{\mu \tilde{r} - \frac{2 \pi t}{\beta}}}{\tilde{r}^{n}}
\end{equation}
The rate at which the particle propagates towards the horizon is the rate at which the size of the applied operator grows. Thus, from \eqref{eqn: u(r,t)}, the butterfly velocity is given as.
\begin{equation}
    \label{eqn: vb}
    v_{B} = \frac{2 \pi }{\beta \mu}
\end{equation}
To determine $v_{B}$, we need to determine $\mu$. To determine $\mu$, we substitute the ansatz \eqref{eqn: ansatz} in \eqref{eqn: ODE for u}. Dropping higher order terms in 1/$\tilde{r}$, we get the following expression for $\mu$. 
\begin{equation}
    \label{eqn: mu expression}
    \mu^{2} = \frac{f_{o}'}{2}(d-1-\theta)
\end{equation}
Substituting \eqref{eqn: mu expression} and \eqref{eqn:temp_k=0} in \eqref{eqn: vb}, we get the following expression for the butterfly velocity.

\begin{equation}
    \label{eqn: vb_EW}
     v_{B}^{2} = \frac{d-1 + z - \theta + q^2 (3 - z + \theta-d)}{2 (d-1 - \theta)}
\end{equation}
The above result matches the result obtained via the shockwave analysis \eqref{eqn:VB_RH=1}.  

\section{$v_{B}$ in terms of the boundary thermodynamic variables}
\label{sec: vb_bound_quant}
The next goal is to understand how boundary thermodynamics and theory parameters ($\theta$ and z) affect information scrambling in HVL theories. In order to achieve that goal, we first express $v_{B}$  (given by \eqref{eqn: vB_HVL}) purely in terms of boundary thermodynamic variables, using a recently developed thermodynamic dictionary for HVL theories \cite{Cong:2024pvs}. This dictionary has various merits, such as the matching of the Smarr formula on the bulk side with the Euler relation on the boundary side, as well as the matching of the first laws on the boundary and in the bulk. Here, we list the dictionary entries that are relavant for expressing $v_B$ in terms of the boundary thermodynamics. The entries for central charge, entropy, and charge  are as follows 
\begin{equation}
    \label{eqn:boundary_central_charge}
    C = \frac{\Omega_{k,d-1}L^{d-\theta-1} r_{F}^{\theta}}{16\pi G_{N}}
\end{equation}
\begin{equation}
    \label{eqn:boundary_entropy}
    S = 4 \pi C  x^{d-\theta -1}
\end{equation}
\begin{equation}
    \label{eqn:boundary_charge}
    Q = \sqrt{2Z_{o}} C q \hspace{0.1cm}[(d-\theta-1)(d-\theta+z -3)]^{\frac{1}{2}}
\end{equation}
Here, x is a bulk parameter given as $x \equiv \frac{r_{h}}{L}$ and S and Q are dimensionless. Note that, in none of the relavant entries (for our purposes), is there an appearance of the arbitrary length parameter $r_{o}$ (refer \cite{Cong:2024pvs}) and thus there is no effect of $r_{o}$ on how information scrambles in the boundary. We continue working with $L=1$ (and $k=0$). $r_{h}$ and q are the two bulk parameters that we need to eliminate in \eqref{eqn: vB_HVL} and thus we express these parameters in terms of the boundary quantities in the following way.
\begin{equation}
    \label{eqn: rh_S/C}
    r_{h} = \tilde{S}^{\frac{1}{d-\theta -1}}
\end{equation}
\begin{equation}
    \label{eqn: q_Q/C}
    q = \frac{\tilde{Q}}{[(d-\theta-1)(d-\theta +z -3)]^{1/2}}
\end{equation}
where $\tilde{S}$ and $\tilde{Q}$ are defined as 
\begin{equation}
    \label{eqn:Stilde}
    \tilde{S} \equiv \frac{S}{4 \pi C}
\end{equation}
\begin{equation}
    \label{eqn: Qtilde}
    \tilde{Q} \equiv \frac{Q}{\sqrt{2 Z_{o}}C}
\end{equation}
Substituting \eqref{eqn: rh_S/C} and \eqref{eqn: q_Q/C} in \eqref{eqn: vB_HVL}, we get the following expression for $v_{B}$ in terms of the boundary thermodynamic variables.

\begin{equation}
    \label{vb_bound_thermo}
    v_{B}^{2} = \frac{\tilde{S}^{\frac{2(z-1)}{d-\theta -1}}(d-1+z-\theta)(d-\theta-1) -\left(\tilde{Q}/\tilde{S}\right)^{2}}{2(d-\theta-1)^{2}}
\end{equation}
We treat $Z_{o}$ as a constant parameter. It is interesting to note that the entropy couples to both z and $\theta$, however charge does not couple to z. With the expression \eqref{vb_bound_thermo} in hand, we now study the variation of $v_B$ with respect to z, $\theta$, $\tilde{S}$ and $\tilde{Q}$. 

\section{Variation of $v_{B}$ with z, $\theta$, $\tilde{S}$ and $\tilde{Q}$}
\label{sec: variation}

In this section, we study the variation of $v_{B}$ and note some interesting features. Before proceeding, there are certain important points to highlight. Given that the Lorentz invariance is broken in HVL theories, $v_B$ is not bounded by $v_B^{S}$ \cite{Mezei:2016wfz}, where $v_B^{S}$ is the butterfly velocity for AdS-Schwarzschild black brane, given as
\begin{equation}
    \label{eqn: vb_schw}
    v_{B}^{S} = \sqrt{\frac{d}{2(d-1)}}
\end{equation}
and is, moreover, super-luminal for certain values of the parameters. Secondly, in this section, we only deal with the "permissible" values of the parameters and thermodynamic variables, where by permissible we mean the combination \{z, $\theta$, $\tilde{S}$, $\tilde{Q}$ \} for which the temperature $T > 0$. Furthermore, we also keep track of the constraints on z and $\theta$ imposed by the null energy condition, discussed in \ref{subsec:constraints}. For analysis in this section, we fix $d=4$. 

First, to analyze only the effect of z, we set $\theta = 0$ (i.e. purely Lifshitz theories). Initially, we deal with the case where $\tilde{Q} =0$ and then consider the effect of non-zero charge. For $\theta = 0$ and $\tilde{Q} = 0$ , 
\begin{figure}[htp]
    \centering
    \begin{subfigure}[t]{0.43\textwidth}
        \centering
        \includegraphics[width=\textwidth]{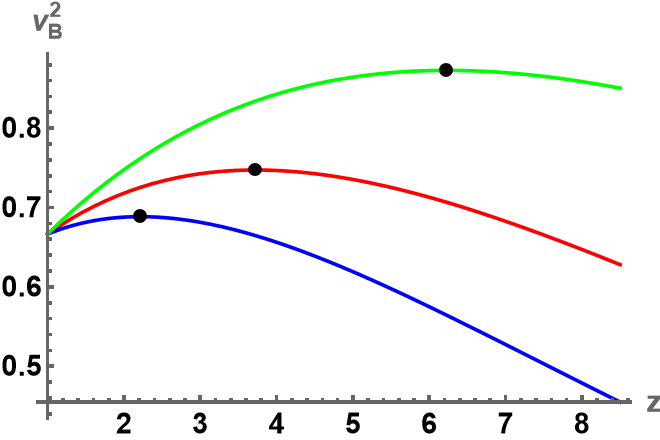} 
        \caption{$\tilde{S} < 1$, $\tilde{Q} = 0$, $\theta = 0$, and $d=4$ for this plot. The blue, red and green curves correspond to $\tilde{S} = 0.7$, $\tilde{S} = 0.75$ and $\tilde{S} = 0.8$ respectively. The black dots represent the non-monotonicities.}
    \end{subfigure}
    \hfill
    \begin{subfigure}[t]{0.43\textwidth}
        \centering
        \includegraphics[width=\textwidth]{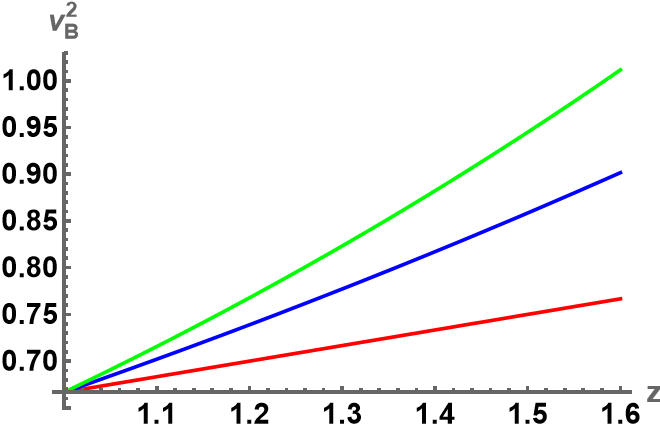} 
        \caption{$\tilde{S} \geq 1$, $\tilde{Q} = 0$, $\theta = 0$ , and $d=4$ for this plot. The red, blue and green curves correspond to $\tilde{S} = 1$, $\tilde{S} = 1.5$ and $\tilde{S} = 2$ respectively.}
    \end{subfigure}
    \caption{$v_B^{2}$ vs z}
    \vspace{0.4em} 

    \begin{subfigure}[t]{0.43\textwidth}
        \centering
        \includegraphics[width=\textwidth]{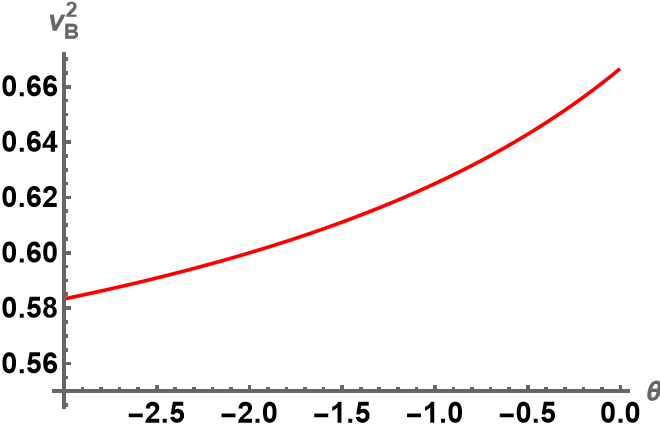} 
        \caption{$\tilde{Q} = 0$, $z =1$, and $d=4$ for this plot. The plot is the same for all values of $\tilde{S}$.}
    \end{subfigure}
    \hfill
    \begin{subfigure}[t]{0.43\textwidth}
        \centering
        \includegraphics[width=\textwidth]{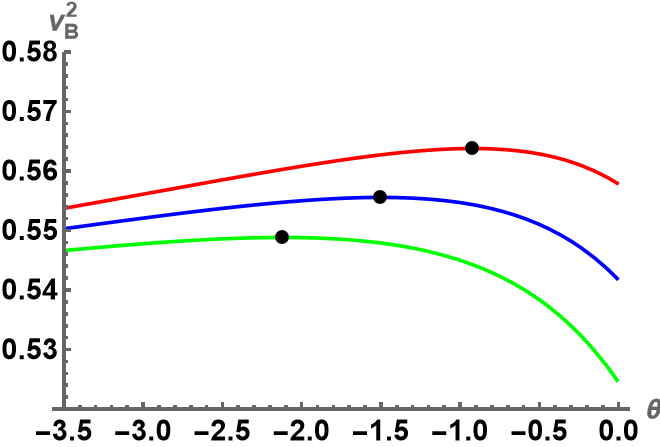} 
        \caption{$\tilde{Q} \neq 0$, $z=1$, $\tilde{S} =1$ and $d=4$ for this plot. The red, blue, and green curves correspond to $\tilde{Q} = 1.4 $, $\tilde{Q} = 1.5 $ and $\tilde{Q} = 1.6 $. The black dots represent the non-monotonicities.}
    \end{subfigure}
    \caption{$v_B^{2} $ vs $\theta$}
    \vspace{0.4em}

    \begin{subfigure}[t]{0.43\textwidth}
        \centering
        \includegraphics[width=\textwidth]{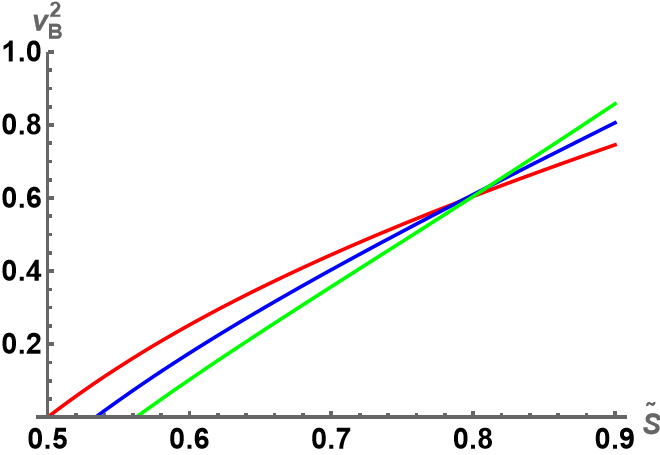} 
        \caption{$\tilde{Q} =1$, $\theta =1$, and $d=4$ for this plot. The green, blue and the red curves correspond to $z =3$, $z = 2.5$, and $z =2$ respectively.}
    \end{subfigure}
    \hfill
    \begin{subfigure}[t]{0.43\textwidth}
        \centering
        \includegraphics[width=\textwidth]{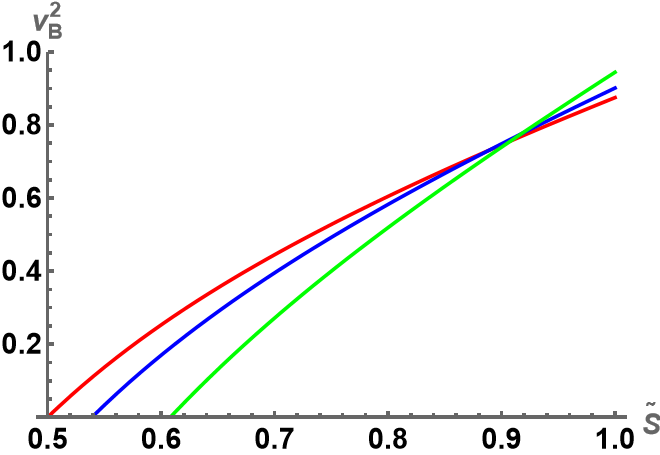} 
        \caption{$\tilde{Q} =1$, $z =2$, and $d=4$ for this plot. The green, blue and the red curves correspond to $\theta =1.5$, $\theta = 1.2$, and $\theta =1$ respectively.}
    \end{subfigure}
    \caption{$v_B^{2}$ vs $\tilde{S}$}    
\end{figure}
\begin{figure}[htp]
    \begin{subfigure}[t]{0.43\textwidth}
        \centering
        \includegraphics[width=\textwidth]{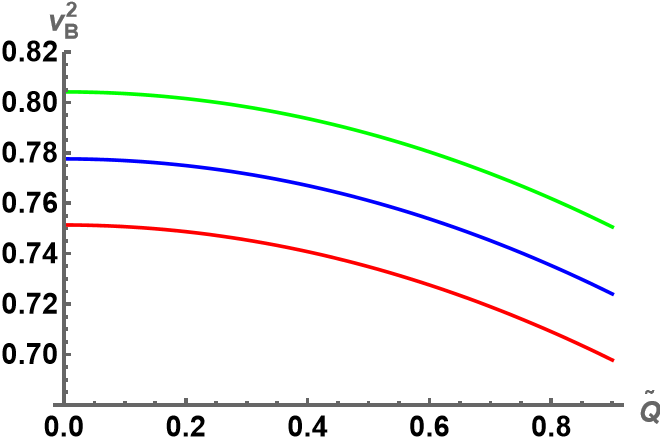} 
        \caption{$\tilde{S} = 1.1$, $\theta = 0.5$, and $d=4$ for this plot. The red, blue and green curves correspond to $z=1.2$, $z=1.3$, and $z=1.4$ respectively. }
    \end{subfigure}
    \hfill
    \begin{subfigure}[t]{0.43\textwidth}
        \centering
        \includegraphics[width=\textwidth]{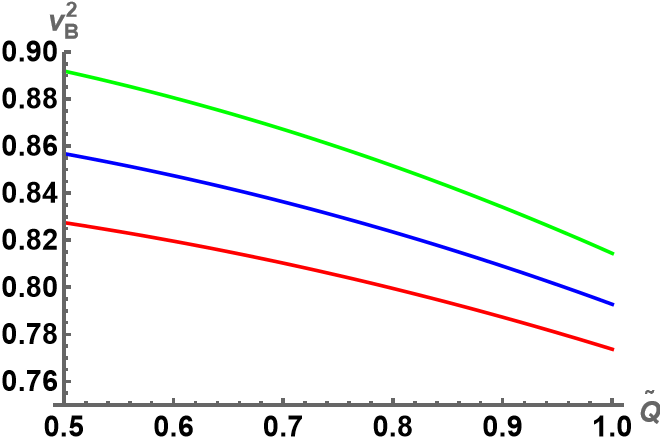} 
        \caption{$\tilde{S} = 1.1$, $z = 1.5$, and $d=4$ for this plot. The red, blue and green curves correspond to $\theta= 0.6 $, $\theta= 0.8 $, and $\theta =1 $ respectively. }
    \end{subfigure}
    \caption{$v_B^{2}$ vs $\tilde{Q}$}
\end{figure}
$v_B$ interestingly exhibits non-monotonic behavior with respect to z when $0 < \tilde{S} < 1$, in the permissible region. We get a maxima. However, only in the range $0.688 < \tilde{S} < 0.877 $, the non-monotonicities lie in the allowed region for z, i.e. ($ z>1 $) and $v_B$ at the non-monotonic point is not superluminal. Since the charge does not couple to z, there is no non-trivial effect of non-zero $\tilde{Q}$. 
$\tilde{Q} \neq 0 $ merely shifts the values and does not change the behavior of $v_{B}$ with respect to z. For $\tilde{S} \geq 1$, $v_{B}$ monotonically increases with z in the permissible region. This suggests that $v_B$ does not necessarily increase with z in Lifshitz theories, as reported in earlier works (such as \cite{Baishya:2024sym}), but it depends on entropy per central charge.

Next, to study the effect of $\theta$, we set $z =1$ (i.e. purely hyperscaling-violating theories). For $z=1$, the corresponding constraint on $\theta$ is $\theta \leq 0 $. We divide the discussion into two subcases $\tilde{Q} = 0$ and $\tilde{Q} \neq 0 $. For $z =1$ and $\tilde{Q} = 0$, $v_{B}$ monotonically increases with $\theta$ for all values of $\tilde{S}$. However, since charge couples with $\theta$, the effect of non-zero charge is non-trivial. $\tilde{Q} \neq 0 $ induces a non-monotonicity in $v_B$ when varied with respect to $\theta.$ We get a maxima (again, we mean in the permissible region, for allowed values of $\theta$ with $v_B$ not being superluminal).  

It is important to note that the non-monotonicities in $v_B$, as a function of $z$ and $\theta$, do not coincide with those in temperature. This can be verified by plotting $T$ and $v_B$ on the same graph. Hence, the underlying causes of the non-monotonicities in $v_B$ and temperature are not the same. 

With respect to $\tilde{S}$, $v_B$ increases monotonically in the permissible region and for all allowed values of $z$ and $\theta$. Note that the entropy dependence of $v_B$ arises due to the appearance of $r_{h}$ in \eqref{eqn: VBOTOC}. However, $v_B$ is a model-dependent parameter, and in certain holographic models, $r_h$ does not appear in the expression for $v_B$. In such models, entropy does not affect information scrambling. One example of this is the planar RNAdS black hole.

On the other hand, $v_B$ monotonically decreases with respect to $\tilde{Q}$ (by which we mean the magnitude of the charge) in the permissible region and for all allowed values of z and $\theta$. It is interesting to note that, in one of the author's upcoming works, quantum chaos in a completely different class of holographic models has been studied, and it is found that $v_B$ necessarily depends on charge and moreover, in all the models studied in the work, $v_B$ is found to monotonically decrease with charge (or more precisely, the magnitude of the charge). This observation suggests that the slowing down of the rate of information scrambling by charge might be a universal feature of all holographic models or even a characteristic of all quantum systems in general.

\section{Discussion and future outlook}
\label{sec: conclusion}
In this work, we presented a detailed discussion on the shockwave analysis in HVL theories and computed various chaos parameters. $v_{B}$ was also computed using the entanglement wedge method, and the results obtained from both methods were found to match. Note that our analysis was restricted to planar horizon topology. One can extend the analysis to hyperbolic and spherical topologies and analyze what effect a non-zero $k$ has on the behavior of $v_B$. However, this task would require devising certain new methods.

Furthermore, using a recently developed thermodynamic dictionary, we expressed $v_B$ purely in terms of boundary thermodynamics, which yielded a neat result. We then discussed the variations of $v_B$ with respect to $z$, $\theta$, $\tilde{S}$, and $\tilde{Q}$. Notably, we observed a non-monotonic behavior of $v_B$ when varied with respect to $z$ for $\tilde{S} < 1$, whereas $v_B$ was found to increase monotonically with $z$ for $\tilde{S} \geq 1$. This implies that $z$ does not necessarily increase $v_B$ (as reported in previous works \cite{Baishya:2024sym}). This suggests that entropy non-trivially affects the rate of information scrambling. Thus, to gain deeper insights into the nature of information scrambling in any holographic model, it is more useful to retain $r_h$ in the expression of $v_B$ (instead of setting it equal to one, as is commonly practiced in the literature).

In uncharged hyperscaling-violating theories, $v_B$ was found to increase monotonically with $\theta$, whereas for non-zero charge, a non-monotonic behavior was obtained. Additionally, $v_B$ was observed to increase monotonically with $\tilde{S}$ and decrease monotonically with $\tilde{Q}$. The latter fact also holds for a completely different class of holographic models discussed in one of the author's upcoming works, and thus one can conjecture that charge decreases the rate of information scrambling in all holographic models or quantum systems in general. It would be interesting to investigate whether a counter-example to this conjecture exists.

When varying $z$ or $\theta$, one navigates through a space of theories. The observed non-monotonicities suggest that the theories on either side of the non-monotonic points may possess distinguishing features, leading to different natures of information scrambling on either side. Unpacking the underlying physics behind these non-monotonicities (as well as other features discussed) is an interesting direction for future work. Additionally, we noted that $v_B$ equals the rate of entanglement growth during thermalization, indicating a connection between the two, which would also be an interesting avenue to explore further.

\section{Acknowledgements}
NL would like to sincerely thank Subhash Mahapatra for his invaluable guidance and fruitful discussions,  Siddhi Swarup Jena for helping with the two figures and Bhaskar Shukla for important discussions. NL would also like to thank Michael Blake, Victor Jahnke, 
Mohammad Reza Mohammadi Mozaffar, Douglas Stanford, and Marika Taylor for their useful comments and suggestions. 

\bibliography{refs.bib}

\providecommand{\href}[2]{#2}\begingroup\raggedright\begin{thebibliography}{10}

\bibitem{Srednicki:1994mfb}
M.~Srednicki, \emph{{Chaos and Quantum Thermalization}}, \href{https://doi.org/10.1103/PhysRevE.50.888}{\emph{Phys. Rev. E} {\bfseries 50} (1994) } [\href{https://arxiv.org/abs/cond-mat/9403051}{{\ttfamily cond-mat/9403051}}].

\bibitem{Suzuki:1996gm}
S.~Suzuki and K.-i.~Maeda, \emph{{Chaos in Schwarzschild space-time: The motion of a spinning particle}}, \href{https://doi.org/10.1103/PhysRevD.55.4848}{\emph{Phys. Rev. D} {\bfseries 55} (1997) 4848} [\href{https://arxiv.org/abs/gr-qc/9604020}{{\ttfamily gr-qc/9604020}}].

\bibitem{Ullmo2012INTRODUCTIONTQ}
D.~Ullmo and S.~Tomsovic, \emph{Introduction to quantum chaos},  2012, \href{https://api.semanticscholar.org/CorpusID:49572714}{https://api.semanticscholar.org/CorpusID:49572714}.

\bibitem{1969JETP...28.1200L}
A.I.~{Larkin} and Y.N.~{Ovchinnikov}, \emph{{Quasiclassical Method in the Theory of Superconductivity}}, {\emph{Soviet Journal of Experimental and Theoretical Physics} {\bfseries 28} (1969) 1200}.

\bibitem{Almheiri:2013hfa}
A.~Almheiri, D.~Marolf, J.~Polchinski, D.~Stanford and J.~Sully, \emph{{An Apologia for Firewalls}}, \href{https://doi.org/10.1007/JHEP09(2013)018}{\emph{JHEP} {\bfseries 09} (2013) 018} [\href{https://arxiv.org/abs/1304.6483}{{\ttfamily 1304.6483}}].

\bibitem{tHooft:1993dmi}
G.~'t~Hooft, \emph{{Dimensional reduction in quantum gravity}}, {\emph{Conf. Proc. C} {\bfseries 930308} (1993) 284} [\href{https://arxiv.org/abs/gr-qc/9310026}{{\ttfamily gr-qc/9310026}}].

\bibitem{Susskind:1994vu}
L.~Susskind, \emph{{The World as a hologram}}, \href{https://doi.org/10.1063/1.531249}{\emph{J. Math. Phys.} {\bfseries 36} (1995) 6377} [\href{https://arxiv.org/abs/hep-th/9409089}{{\ttfamily hep-th/9409089}}].

\bibitem{Maldacena:1997re}
J.M.~Maldacena, \emph{{The Large N limit of superconformal field theories and supergravity}}, \href{https://doi.org/10.4310/ATMP.1998.v2.n2.a1}{\emph{Adv. Theor. Math. Phys.} {\bfseries 2} (1998) 231} [\href{https://arxiv.org/abs/hep-th/9711200}{{\ttfamily hep-th/9711200}}].

\bibitem{Witten:1998qj}
E.~Witten, \emph{{Anti-de Sitter space and holography}}, \href{https://doi.org/10.4310/ATMP.1998.v2.n2.a2}{\emph{Adv. Theor. Math. Phys.} {\bfseries 2} (1998) 253} [\href{https://arxiv.org/abs/hep-th/9802150}{{\ttfamily hep-th/9802150}}].

\bibitem{Gubser:1998bc}
S.S.~Gubser, I.R.~Klebanov and A.M.~Polyakov, \emph{{Gauge theory correlators from noncritical string theory}}, \href{https://doi.org/10.1016/S0370-2693(98)00377-3}{\emph{Phys. Lett. B} {\bfseries 428} (1998) 105} [\href{https://arxiv.org/abs/hep-th/9802109}{{\ttfamily hep-th/9802109}}].

\bibitem{Kachru_2008}
S.~Kachru, X.~Liu and M.~Mulligan, \emph{Gravity duals of lifshitz-like fixed points}, \href{https://doi.org/10.1103/physrevd.78.106005}{\emph{Physical Review D} {\bfseries 78} (2008) }.

\bibitem{Balasubramanian:2008dm}
K.~Balasubramanian and J.~McGreevy, \emph{{Gravity duals for non-relativistic CFTs}}, \href{https://doi.org/10.1103/PhysRevLett.101.061601}{\emph{Phys. Rev. Lett.} {\bfseries 101} (2008) 061601} [\href{https://arxiv.org/abs/0804.4053}{{\ttfamily 0804.4053}}].

\bibitem{Taylor:2008tg}
M.~Taylor, \emph{{Non-relativistic holography}},  \href{https://arxiv.org/abs/0812.0530}{{\ttfamily 0812.0530}}.

\bibitem{Ayon-Beato}
E.~Ayón-Beato, A.~Garbarz, G.~Giribet and M.~Hassaïne, \emph{Lifshitz black hole in three dimensions}, \href{https://doi.org/10.1103/PhysRevD.80.104029}{\emph{Physical Review D} {\bfseries 80} (2009) }.

\bibitem{Mann:2009yx}
R.B.~Mann, \emph{{Lifshitz Topological Black Holes}}, \href{https://doi.org/10.1088/1126-6708/2009/06/075}{\emph{JHEP} {\bfseries 06} (2009) 075} [\href{https://arxiv.org/abs/0905.1136}{{\ttfamily 0905.1136}}].

\bibitem{Bertoldi:2009vn}
G.~Bertoldi, B.A.~Burrington and A.~Peet, \emph{{Black Holes in asymptotically Lifshitz spacetimes with arbitrary critical exponent}}, \href{https://doi.org/10.1103/PhysRevD.80.126003}{\emph{Phys. Rev. D} {\bfseries 80} (2009) 126003} [\href{https://arxiv.org/abs/0905.3183}{{\ttfamily 0905.3183}}].

\bibitem{Widom1965SurfaceTA}
B.~Widom, \emph{Surface tension and molecular correlations near the critical point}, {\emph{Journal of Chemical Physics} {\bfseries 43} (1965) 3892}.

\bibitem{Gouteraux:2011ce}
B.~Gouteraux and E.~Kiritsis, \emph{{Generalized Holographic Quantum Criticality at Finite Density}}, \href{https://doi.org/10.1007/JHEP12(2011)036}{\emph{JHEP} {\bfseries 12} (2011) 036} [\href{https://arxiv.org/abs/1107.2116}{{\ttfamily 1107.2116}}].

\bibitem{Huijse:2011ef}
L.~Huijse, S.~Sachdev and B.~Swingle, \emph{{Hidden Fermi surfaces in compressible states of gauge-gravity duality}}, \href{https://doi.org/10.1103/PhysRevB.85.035121}{\emph{Phys. Rev. B} {\bfseries 85} (2012) 035121} [\href{https://arxiv.org/abs/1112.0573}{{\ttfamily 1112.0573}}].

\bibitem{Alishahiha:2012qu}
M.~Alishahiha, E.~O~Colgain and H.~Yavartanoo, \emph{{Charged Black Branes with Hyperscaling Violating Factor}}, \href{https://doi.org/10.1007/JHEP11(2012)137}{\emph{JHEP} {\bfseries 11} (2012) 137} [\href{https://arxiv.org/abs/1209.3946}{{\ttfamily 1209.3946}}].

\bibitem{Dong:2012se}
X.~Dong, S.~Harrison, S.~Kachru, G.~Torroba and H.~Wang, \emph{{Aspects of holography for theories with hyperscaling violation}}, \href{https://doi.org/10.1007/JHEP06(2012)041}{\emph{JHEP} {\bfseries 06} (2012) 041} [\href{https://arxiv.org/abs/1201.1905}{{\ttfamily 1201.1905}}].

\bibitem{Gouteraux:2012yr}
B.~Gouteraux and E.~Kiritsis, \emph{{Quantum critical lines in holographic phases with (un)broken symmetry}}, \href{https://doi.org/10.1007/JHEP04(2013)053}{\emph{JHEP} {\bfseries 04} (2013) 053} [\href{https://arxiv.org/abs/1212.2625}{{\ttfamily 1212.2625}}].

\bibitem{Gath}
J.~Gath, J.~Hartong, R.~Monteiro and N.~Obers, \emph{Holographic models for theories with hyperscaling violation}, \href{https://doi.org/10.1007/JHEP04(2013)159}{\emph{Journal of High Energy Physics} {\bfseries 2013} (2012) }.

\bibitem{Bueno}
P.~Bueno, W.~Chemissany and C.~Shahbazi, \emph{On hvlif-like solutions in gauged supergravity}, \href{https://doi.org/10.1140/epjc/s10052-013-2684-3}{\emph{The European Physical Journal C} {\bfseries 74} (2012) }.

\bibitem{Pedraza:2018eey}
J.F.~Pedraza, W.~Sybesma and M.R.~Visser, \emph{{Hyperscaling violating black holes with spherical and hyperbolic horizons}}, \href{https://doi.org/10.1088/1361-6382/ab0094}{\emph{Class. Quant. Grav.} {\bfseries 36} (2019) 054002} [\href{https://arxiv.org/abs/1807.09770}{{\ttfamily 1807.09770}}].

\bibitem{Shenker_2014}
S.H.~Shenker and D.~Stanford, \emph{Black holes and the butterfly effect}, \href{https://doi.org/10.1007/jhep03(2014)067}{\emph{Journal of High Energy Physics} {\bfseries 2014} (2014) }.

\bibitem{Roberts_2015}
D.A.~Roberts, D.~Stanford and L.~Susskind, \emph{Localized shocks}, \href{https://doi.org/10.1007/jhep03(2015)051}{\emph{Journal of High Energy Physics} {\bfseries 2015} (2015) }.

\bibitem{PhysRevLett.117.091602}
D.A.~Roberts and B.~Swingle, \emph{Lieb-robinson bound and the butterfly effect in quantum field theories}, \href{https://doi.org/10.1103/PhysRevLett.117.091602}{\emph{Phys. Rev. Lett.} {\bfseries 117} (2016) 091602}.

\bibitem{Perlmutter:2016pkf}
E.~Perlmutter, \emph{{Bounding the Space of Holographic CFTs with Chaos}}, \href{https://doi.org/10.1007/JHEP10(2016)069}{\emph{JHEP} {\bfseries 10} (2016) 069} [\href{https://arxiv.org/abs/1602.08272}{{\ttfamily 1602.08272}}].

\bibitem{Jahnke:2018off}
V.~Jahnke, \emph{{Recent developments in the holographic description of quantum chaos}}, \href{https://doi.org/10.1155/2019/9632708}{\emph{Adv. High Energy Phys.} {\bfseries 2019} (2019) 9632708} [\href{https://arxiv.org/abs/1811.06949}{{\ttfamily 1811.06949}}].

\bibitem{Sekino_2008}
Y.~Sekino and L.~Susskind, \emph{Fast scramblers}, \href{https://doi.org/10.1088/1126-6708/2008/10/065}{\emph{Journal of High Energy Physics} {\bfseries 2008} (2008) 065–065}.

\bibitem{Maldacena:2001kr}
J.M.~Maldacena, \emph{{Eternal black holes in anti-de Sitter}}, \href{https://doi.org/10.1088/1126-6708/2003/04/021}{\emph{JHEP} {\bfseries 04} (2003) 021} [\href{https://arxiv.org/abs/hep-th/0106112}{{\ttfamily hep-th/0106112}}].

\bibitem{Shenker:2014cwa}
S.H.~Shenker and D.~Stanford, \emph{{Stringy effects in scrambling}}, \href{https://doi.org/10.1007/JHEP05(2015)132}{\emph{JHEP} {\bfseries 05} (2015) 132} [\href{https://arxiv.org/abs/1412.6087}{{\ttfamily 1412.6087}}].

\bibitem{PhysRevA.94.040302}
B.~Swingle, G.~Bentsen, M.~Schleier-Smith and P.~Hayden, \emph{Measuring the scrambling of quantum information}, \href{https://doi.org/10.1103/PhysRevA.94.040302}{\emph{Phys. Rev. A} {\bfseries 94} (2016) 040302}.

\bibitem{PhysRevA.94.062329}
G.~Zhu, M.~Hafezi and T.~Grover, \emph{Measurement of many-body chaos using a quantum clock}, \href{https://doi.org/10.1103/PhysRevA.94.062329}{\emph{Phys. Rev. A} {\bfseries 94} (2016) 062329}.

\bibitem{PhysRevX.7.031011}
J.~Li, R.~Fan, H.~Wang, B.~Ye, B.~Zeng, H.~Zhai et~al., \emph{Measuring out-of-time-order correlators on a nuclear magnetic resonance quantum simulator}, \href{https://doi.org/10.1103/PhysRevX.7.031011}{\emph{Phys. Rev. X} {\bfseries 7} (2017) 031011}.

\bibitem{Yao:2016ayk}
N.Y.~Yao, F.~Grusdt, B.~Swingle, M.D.~Lukin, D.M.~Stamper-Kurn, J.E.~Moore et~al., \emph{{Interferometric Approach to Probing Fast Scrambling}},  \href{https://arxiv.org/abs/1607.01801}{{\ttfamily 1607.01801}}.

\bibitem{Cong:2024pvs}
W.~Cong, D.~Kubiz\v{n}\'ak, R.B.~Mann and M.R.~Visser, \emph{{Holographic dictionary for Lifshitz and hyperscaling violating black holes}},  \href{https://arxiv.org/abs/2410.16145}{{\ttfamily 2410.16145}}.

\bibitem{Baishya:2024sym}
B.~Baishya, A.~Chakraborty and N.~Padhi, \emph{{A study of three butterflies: entanglement wedge method, OTOC and pole-skipping}},  \href{https://arxiv.org/abs/2406.18319}{{\ttfamily 2406.18319}}.

\bibitem{Alishahiha:2014cwa}
M.~Alishahiha, A.~Faraji~Astaneh and M.R.~Mohammadi~Mozaffar, \emph{{Thermalization in backgrounds with hyperscaling violating factor}}, \href{https://doi.org/10.1103/PhysRevD.90.046004}{\emph{Phys. Rev. D} {\bfseries 90} (2014) 046004} [\href{https://arxiv.org/abs/1401.2807}{{\ttfamily 1401.2807}}].

\bibitem{Czech:2012bh}
B.~Czech, J.L.~Karczmarek, F.~Nogueira and M.~Van~Raamsdonk, \emph{{The Gravity Dual of a Density Matrix}}, \href{https://doi.org/10.1088/0264-9381/29/15/155009}{\emph{Class. Quant. Grav.} {\bfseries 29} (2012) 155009} [\href{https://arxiv.org/abs/1204.1330}{{\ttfamily 1204.1330}}].

\bibitem{Wall:2012uf}
A.C.~Wall, \emph{{Maximin Surfaces, and the Strong Subadditivity of the Covariant Holographic Entanglement Entropy}}, \href{https://doi.org/10.1088/0264-9381/31/22/225007}{\emph{Class. Quant. Grav.} {\bfseries 31} (2014) 225007} [\href{https://arxiv.org/abs/1211.3494}{{\ttfamily 1211.3494}}].

\bibitem{Headrick:2014cta}
M.~Headrick, V.E.~Hubeny, A.~Lawrence and M.~Rangamani, \emph{{Causality \& holographic entanglement entropy}}, \href{https://doi.org/10.1007/JHEP12(2014)162}{\emph{JHEP} {\bfseries 12} (2014) 162} [\href{https://arxiv.org/abs/1408.6300}{{\ttfamily 1408.6300}}].

\bibitem{Dong:2016eik}
X.~Dong, D.~Harlow and A.C.~Wall, \emph{{Reconstruction of Bulk Operators within the Entanglement Wedge in Gauge-Gravity Duality}}, \href{https://doi.org/10.1103/PhysRevLett.117.021601}{\emph{Phys. Rev. Lett.} {\bfseries 117} (2016) 021601} [\href{https://arxiv.org/abs/1601.05416}{{\ttfamily 1601.05416}}].

\bibitem{Mezei:2016wfz}
M.~Mezei and D.~Stanford, \emph{{On entanglement spreading in chaotic systems}}, \href{https://doi.org/10.1007/JHEP05(2017)065}{\emph{JHEP} {\bfseries 05} (2017) 065} [\href{https://arxiv.org/abs/1608.05101}{{\ttfamily 1608.05101}}].

\bibitem{Ryu_2006}
S.~Ryu and T.~Takayanagi, \emph{Holographic derivation of entanglement entropy from the anti–de sitter space/conformal field theory correspondence}, \href{https://doi.org/10.1103/physrevlett.96.181602}{\emph{Physical Review Letters} {\bfseries 96} (2006) }.

\bibitem{Hubeny:2007xt}
V.E.~Hubeny, M.~Rangamani and T.~Takayanagi, \emph{{A Covariant holographic entanglement entropy proposal}}, \href{https://doi.org/10.1088/1126-6708/2007/07/062}{\emph{JHEP} {\bfseries 07} (2007) 062} [\href{https://arxiv.org/abs/0705.0016}{{\ttfamily 0705.0016}}].

\bibitem{Dong_2022}
X.~Dong, D.~Wang, W.W.~Weng and C.-H.~Wu, \emph{A tale of two butterflies: an exact equivalence in higher-derivative gravity}, \href{https://doi.org/10.1007/jhep10(2022)009}{\emph{Journal of High Energy Physics} {\bfseries 2022} (2022) }.

\end{thebibliography}\endgroup
\bibliographystyle{JHEP}

\end{document}